\documentclass[conference]{IEEEtran}
\IEEEoverridecommandlockouts
\usepackage{cite}
\usepackage{amsmath,amssymb,amsfonts}
\usepackage{algorithmic}
\usepackage{graphicx}
\usepackage{textcomp}
\usepackage{xcolor}
\usepackage{array}
\usepackage{booktabs}
\usepackage{float}
\usepackage{todonotes}
\usepackage{parskip}

\def\BibTeX{{\rm B\kern-.05em{\sc i\kern-.025em b}\kern-.08em
    T\kern-.1667em\lower.7ex\hbox{E}\kern-.125emX}}

\begin{document}

\title{Improving Early Detection of Scientifically Interesting Radio Sources Using Enhanced Isolation Forest\\
\thanks{ISBN 979-8-3195-1782-1/26/\$31.00~\textcopyright{}2026 IEEE}}

\author{
\IEEEauthorblockN{Pfunzo Thavhana}
\IEEEauthorblockA{
\textit{Department of Computer Science} \\
\textit{University of the Western Cape} \\
Cape Town, South Africa\\
3805564@myuwc.ac.za}
\and
% \IEEEauthorblockN{Clement Nyirenda}
% \IEEEauthorblockA{
% University of the Western Cape\\
% Cape Town, South Africa\\
% cnyirenda@uwc.ac.za}

\and
\IEEEauthorblockN{Clement Nyirenda}
\IEEEauthorblockA{
\textit{Department of Computer Science}\\
\textit{eResearch Office}\\
\textit{University of the Western Cape}\\
Cape Town, South Africa\\
cnyirenda@uwc.ac.za}

\and
\IEEEauthorblockN{Michelle Lochner}
\IEEEauthorblockA{
University of the Western Cape\\
\textit{Department of Physics and Astronomy}\\
Cape Town, South Africa\\
mlochner@uwc.ac.za}
}

\maketitle

\begin{abstract}
The rapid growth of radio astronomical surveys has created a major challenge in identifying scientifically interesting sources from large populations of observed objects. This study investigates whether enhanced Isolation Forest (iForest) methods can improve the early detection and ranking of scientifically interesting radio sources. Four iForest variants, Standard Enhanced, Pooled Gain, Kurtosis-based, and Hypervolume, are evaluated using 1,031 extended radio sources from the MeerKAT Galaxy Cluster Legacy Survey (MGCLS). Each source is represented by a 52-dimensional feature set derived using Bootstrap Your Own Latent (BYOL) self-supervised learning and principal component analysis (PCA), while expert scientific-interest ratings are used to define the ground-truth labels. The methods are evaluated using Average Precision (AP), receiver operating characteristic area under the curve (ROC AUC), Precision@10, Precision@50, and cumulative anomaly detection. Pooled Gain achieves the highest Average Precision (AP = 0.201) and the strongest early-stage ranking performance ($P@10$ = 0.600), identifying six expert-validated anomalies among the ten highest-ranked sources. Standard Enhanced and Kurtosis-based achieve the highest ROC AUC (0.677), while the AUC values of the four variants range from 0.653 to 0.677. These results indicate that feature-aware isolation strategies can improve the prioritization of scientifically interesting sources at the top of the ranked list, although overall discrimination remains comparable across the evaluated variants. The findings support the use of enhanced Isolation Forest methods as an initial ranking mechanism within human-in-the-loop systems for large-scale radio astronomical surveys.
\end{abstract}


\section{Introduction}

\subsection{The Data Deluge in Modern Radio Astronomy}

The Square Kilometre Array (SKA) and its precursors represent a paradigm shift in observational astronomy \cite{b1}. MeerKAT, the SKA precursor in South Africa, generates approximately 2 terabytes of data per hour during typical observations \cite{b2}. The MeerKAT Galaxy Cluster Legacy Survey (MGCLS) alone has cataloged over 1 million radio sources, each requiring classification and analysis \cite{b2}. At a conservative estimate of 30 seconds per source for human inspection, complete manual review would require over 8,300 person-hours, an impossibility for any research team \cite{b3}. This data avalanche requires automated methods to identify the approximately 1\% of sources that exhibit scientifically interesting properties while filtering out the remaining 99\% that display standard morphologies \cite{b3,b4}. Previous work by Lochner and Bassett \cite{b3} demonstrated that machine-learning approaches, particularly anomaly-detection algorithms, can effectively address this challenge when combined with human expertise.

\subsection{Astrophysical Anomalies in Radio Astronomy}

Astrophysical anomalies are sources with unusual morphological, spectral,
or temporal characteristics that may be of scientific interest
\cite{b3,b4}. Morphological examples include X-shaped radio galaxies,
whose unusual wings may be associated with backflow or a reorientation
of the jet axis following a merger \cite{b6}; one-sided radio sources,
for which relativistic beaming is a possible explanation \cite{b7};
giant radio galaxies that extend to megaparsec scales and can be used
to probe the intergalactic medium \cite{b8}; and the recently identified
Odd Radio Circles (ORCs) \cite{b10}.

Other examples include transient and variable phenomena such as
millisecond-duration Fast Radio Bursts (FRBs) \cite{b12}, pulsars with
unusual timing behaviour \cite{b13}, and radio supernovae \cite{b14}.
Scientifically interesting radio sources can also include radio phoenixes,
in which aged radio plasma is reactivated by compression or related
mechanisms \cite{b16}, dying radio galaxies with steep-spectrum emission
associated with declining nuclear activity \cite{b15}, and hybrid
morphology radio sources that exhibit different Fanaroff--Riley
morphologies on opposite sides of the same source \cite{b18}.

\subsection{State of the Art and Anomaly Detection Challenge}
The increasing volume and complexity of astronomical survey data have
motivated machine-learning methods for identifying unusual or
scientifically valuable sources. Common approaches include supervised
classification, clustering, representation learning, and unsupervised
anomaly detection. Because rare astronomical phenomena are often poorly
represented in labeled datasets, unsupervised and semi-supervised methods
are particularly attractive for discovering unknown or underrepresented
source populations.

Isolation Forest (iForest) is an efficient unsupervised anomaly-detection
method that identifies observations that are easier to isolate through
recursive partitioning \cite{b23}. However, its random feature and split
selection may be suboptimal when informative dimensions are unevenly
distributed or scientifically interesting sources form structured
populations in feature space \cite{b24}.

Human-in-the-loop frameworks such as ASTRONOMALY and ASTRONOMALY Protege
combine automated anomaly ranking with expert feedback to prioritize
scientifically interesting sources \cite{b3,b4}. Related work in cloud
security, IoT intrusion detection, and mobile edge computing similarly
demonstrates the use of application-specific machine-learning strategies
\cite{b29,b30,b31}, although these objectives differ from astronomical
anomaly discovery. Despite these advances, statistical unusualness does
not necessarily correspond to scientific interest, and the comparative
effectiveness of feature-aware and alternative Isolation Forest scoring
strategies remains unclear \cite{b4,b24,b25}.

% Despite these advances, important limitations remain in astronomical anomaly detection. Existing frameworks such as ASTRONOMALY and ASTRONOMALY Protege demonstrate the value of combining automated discovery with expert feedback, but they also highlight the distinction between statistical unusualness and scientific interest \cite{b24,b4}. A source that is statistically unusual is not necessarily scientifically valuable, while an important source may not be sufficiently isolated within a particular feature space. Consequently, the effectiveness of a human-in-the-loop discovery system depends not only on the quality of its feature representation but also on its ability to rank scientifically interesting candidates near the top of the list.

% Furthermore, although feature-aware and alternative scoring strategies have been proposed for Isolation Forest \cite{b24,b25}, their comparative effectiveness for prioritizing expert-validated scientifically interesting radio sources has not been systematically established. This motivates the investigation of enhanced Isolation Forest strategies that can better exploit the structure of high-dimensional astronomical feature spaces and improve the initial prioritization of candidates for expert inspection.

\subsection{Research Problem and Motivation}

Modern radio astronomical surveys contain very large numbers of sources, making exhaustive visual inspection by domain experts impractical \cite{b3,b4}. Automated anomaly-detection methods can reduce this burden by ranking sources according to their degree of unusualness \cite{b3,b11}. However, a fundamental limitation is that statistical unusualness does not necessarily correspond to scientific interest \cite{b3,b4}. Imaging artefacts, noise-related structures, or sources that are unusual only within a particular feature representation may receive high anomaly scores, while scientifically important sources may not appear as strong statistical outliers \cite{b3,b4}.

%MAYI
% This limitation is particularly relevant to Isolation Forest, where conventional random feature selection may not fully exploit informative dimensions in high-dimensional astronomical feature spaces \cite{b23,b24}. Recent extensions have proposed feature-aware splitting and alternative scoring mechanisms \cite{b24,b25}, but their effectiveness for prioritizing expert-validated scientifically interesting radio sources has not been systematically established. In particular, it remains unclear whether these enhanced strategies provide meaningful advantages when experts have limited time and can inspect only a small number of highly ranked candidates.

This limitation is particularly relevant to Isolation Forest, where conventional random feature selection may not fully exploit informative dimensions in high-dimensional astronomical feature spaces \cite{b23,b24}. Feature-aware splitting and alternative scoring mechanisms have therefore been proposed to improve anomaly ranking \cite{b24,b25}.

Therefore, the central problem addressed in this study is how to improve the early-stage ranking of scientifically interesting radio sources using unsupervised Isolation Forest-based anomaly detection. The study investigates whether enhanced feature-selection and anomaly-scoring strategies can identify a greater concentration of expert-validated anomalies near the top of the ranked candidate list than alternative Isolation Forest variants. Addressing this problem could improve the efficiency of human-in-the-loop discovery systems by allowing astronomers to focus their limited inspection time on a smaller and more informative set of candidate sources.

\subsubsection{Limitations of Existing Astronomical Anomaly Detection}

Existing astronomical anomaly-detection frameworks such as ASTRONOMALY and ASTRONOMALY Protege demonstrate the value of combining automated source ranking with expert feedback \cite{b3, b4}. However, their effectiveness depends on the initial feature representation and anomaly-ranking mechanism, since statistical unusualness does not necessarily correspond to scientific interest. This motivates alternative Isolation Forest strategies that can improve candidate prioritization before human feedback is applied.

%MAYI SUMMARIZED THIS, THERE WAS A LOT OF REPITITION
% Existing astronomical anomaly-detection frameworks have demonstrated that machine learning can substantially reduce the amount of data requiring direct expert inspection. However, the effectiveness of such systems depends strongly on the quality of the underlying feature representation and anomaly-ranking mechanism. ASTRONOMALY provides a human-in-the-loop framework in which expert feedback is used to iteratively refine the selection of interesting sources \cite{b3}, while ASTRONOMALY Protege extends this concept by learning user-defined scientific interest in a self-supervised feature space \cite{b4}.

% Although these approaches demonstrate the value of combining automated discovery with expert feedback, they also highlight an important limitation of conventional anomaly detection: statistical unusualness and scientific interest are not necessarily equivalent \cite{b4}. In particular, if the initial anomaly-ranking mechanism fails to prioritize informative regions of the feature space, scientifically valuable sources may be ranked too low for efficient expert inspection. This motivates investigation of alternative Isolation Forest strategies that can improve the initial ranking of candidates before human feedback is applied.

\subsection{Research Gap}

%MAYI
% Existing astronomical anomaly detection frameworks have demonstrated the potential of automated source ranking and human-in-the-loop feedback to reduce the burden of expert inspection in large astronomical surveys \cite{b3,b4}. However, conventional Isolation Forest relies on randomized feature and split selection, which may not fully exploit informative structure within high-dimensional feature spaces. Recent studies have investigated more informed feature selection and splitting strategies, as well as alternative distribution- and volume-based anomaly scoring mechanisms, demonstrating that modifications to the standard Isolation Forest framework can affect anomaly ranking performance \cite{b24,b25}. Nevertheless, comparatively little attention has been given to evaluating these alternative Isolation Forest strategies specifically for prioritizing sources according to expert-defined scientific interest in radio astronomical data. In particular, it remains important to determine whether such strategies can improve the concentration of scientifically interesting sources near the top of the ranked list, where expert inspection is most limited. This study addresses this gap by systematically comparing four Isolation Forest variants using a common MGCLS feature representation and expert-validated evaluation subset.

Existing astronomical anomaly-detection frameworks demonstrate the potential of automated source ranking and human-in-the-loop feedback \cite{b3,b4}. However, comparatively little attention has been given to evaluating alternative Isolation Forest strategies for prioritizing expert-defined scientific interest in radio astronomical data. In particular, it remains unclear whether feature-aware and alternative scoring strategies can improve the concentration of scientifically interesting sources near the top of the ranked list \cite{b24,b25}. This study addresses this gap by comparing four Isolation Forest variants using a common MGCLS feature representation and expert-validated evaluation subset.

\subsection{Research Objectives and Contributions}

The primary objective of this study is to investigate whether enhanced Isolation Forest strategies can improve the early-stage ranking of scientifically interesting radio sources. To achieve this objective, the study pursues the following specific objectives:

\begin{enumerate}
\item Evaluate four Isolation Forest variants: Standard Enhanced, Pooled Gain, Kurtosis-based, and Hypervolume using a common set of radio astronomical features.
\item Assess the effectiveness of the four variants using expert-validated scientific-interest labels derived from the MGCLS evaluation subset.
\item Compare global and ranking-oriented performance using Average Precision (AP), receiver operating characteristic area under the curve (ROC AUC), Precision@10 (P@10), Precision@50 (P@50), and cumulative anomaly detection.
\item Determine whether feature-aware and alternative scoring strategies improve early-stage prioritization, particularly when only a small number of highly ranked sources can be inspected by experts.
\item Identify the most suitable Isolation Forest strategy for human-in-the-loop source prioritization based on the observed trade-offs between global discrimination and top-ranked candidate quality.
\end{enumerate}

The main contributions of this study are:
\begin{itemize}
\item a systematic empirical comparison of four enhanced Isolation Forest strategies, Standard Enhanced, Pooled Gain, Kurtosis-based, and Hypervolume for anomaly ranking in radio astronomical data;
\item an evaluation framework using expert-validated scientific-interest labels and both global-discrimination and top-ranked retrieval measures;
\item an analysis of early-stage candidate prioritization, reflecting the practical situation in which astronomers can inspect only a limited number of sources;
\item an empirical characterization of the trade-off between overall anomaly discrimination and the quality of the highest-ranked candidates; and
\item a quantitative assessment of the suitability of enhanced Isolation Forest strategies as an initial ranking component for human-in-the-loop discovery of scientifically interesting radio sources.
\end{itemize}

\section{Related Work}

\subsection{The Traditional Isolation Forest Technique}

% Introduced by Liu, Ting, and Zhou in 2008 \cite{b23}, Isolation Forest represents a paradigm shift in anomaly detection philosophy. Unlike most methods that construct profiles of ``normal'' behavior, iForest explicitly exploits the property that anomalies are both ``few and different'' from normal instances. The algorithm's elegance lies in its simplicity: anomalies require fewer random partitions to isolate than normal points. Technically, iForest operates through an ensemble of Isolation Trees (iTrees). Each tree is constructed by recursively partitioning randomly selected subsamples through random feature and split-value selection. The anomaly score for a point is derived from its average path length across all trees, normalized by the expected path length for the dataset size:

Isolation Forest identifies anomalies as observations that require fewer random partitions to isolate than normal instances \cite{b23}. It constructs an ensemble of Isolation Trees (iTrees) using random feature and split-value selection, with anomaly scores derived from the normalized average path length:

{\small
\begin{equation}
s(x,n) = 2^{-\frac{E[h(x)]}{c(n)}}.
\end{equation}
}
{\small
\begin{equation}
c(n) = 2H(n-1)-\frac{2(n-1)}{n},
\end{equation}
}

where $E(h(x))$ is the expected path length and $c(n)$ is a normalization factor \cite{b23}. For an ensemble of $t$ trees trained on subsamples of size $\psi$, the method has approximately $O(t\psi\log\psi)$ training complexity, requires no labeled anomaly data, and is suitable for high-dimensional spaces.

\subsection{Enhanced Isolation Forest Variants}

The standard iForest algorithm's reliance on random splits presents limitations for certain anomaly types. Cortes \cite{b24} systematically investigated these limitations, demonstrating that clustered anomalies could be better detected using feature-aware strategies. The Pooled Gain variant weights feature selection by combining variance (70\%) and kurtosis (30\%):

{\small
\begin{equation}
I(f_i) =
0.7\frac{\mathrm{Var}(f_i)}
{\sum_j \mathrm{Var}(f_j)}
+
0.3\frac{|\mathrm{Kurt}(f_i)|}
{\sum_j |\mathrm{Kurt}(f_j)|},
\end{equation}
}
The feature-selection probability is obtained by normalizing the resulting
importance values:
{\small
\begin{equation}
P(f_i)=\frac{I(f_i)}{\sum_j I(f_j)}.
\end{equation}
}
This weighting scheme increases the probability of selecting features with larger variance and heavier-tailed distributions, potentially improving detection when anomalous structure is concentrated in such dimensions \cite{b24}. The Kurtosis-based variant uses normalized absolute kurtosis alone to guide feature selection toward dimensions with heavier-tailed distributions.
The corresponding feature-selection probability is
{\small
\begin{equation}
P_{K}(f_i)=\frac{|\mathrm{Kurt}(f_i)|}{\sum_j |\mathrm{Kurt}(f_j)|}.
\end{equation}
}

For the Hypervolume variant, the implementation used a leaf-size-based
scoring function. For a source $x$ reaching a terminal leaf containing
$m_T(x)$ observations in tree $T$, the leaf-level score was defined as
{\small
\begin{equation}
v(x,T)=\frac{1}{m_T(x)+1}.
\end{equation}
}
The final anomaly score was obtained by averaging the leaf-level scores
across the $N$ isolation trees:
{\small
\begin{equation}
V(x)=\frac{1}{N}\sum_{T=1}^{N}v(x,T).
\end{equation}
}
where $m_T(x)$ denotes the number of observations in the terminal leaf
reached by source $x$ in tree $T$, and $N=100$ is the number of
isolation trees. Higher values of $V(x)$ indicate greater anomaly
likelihood in the implemented scoring scheme.

\subsection{Human-in-the-Loop Systems in Astronomy}

The ASTRONOMALY framework \cite{b3} combines initial anomaly detection with expert labeling and active learning to refine candidate selection according to scientific interest. ASTRONOMALY Protege \cite{b4} extends this approach through BYOL-based self-supervised feature extraction and active regression, and demonstrated the discovery of scientifically valuable sources in MGCLS, including X-shaped galaxies and filamentary structures. Together, these systems show how automated ranking can reduce the search space while expert feedback provides the scientific context needed to distinguish meaningful anomalies from statistical curiosities.

% The ASTRONOMALY framework \cite{b3} established the standard for human-in-the-loop discovery in astronomy. Its pipeline begins with initial anomaly detection using iForest, proceeds to expert labeling of top candidates, employs active learning to refine the model based on feedback, and iteratively improves through this cycle. This approach allows the system to learn which types of statistical unusualness correspond to scientific interest for specific research goals \cite{b3}.

% Building on this foundation, ASTRONOMALY Protege \cite{b4} introduced critical enhancements: self-supervised deep learning for feature extraction using BYOL, active regression for modeling continuous interest scores, and an improved user interface. Validation on the MGCLS dataset demonstrated that this integrated approach could discover scientifically valuable sources, including X-shaped galaxies and filamentary structures \cite{b4}. The success of these systems underscores the complementary strengths of algorithms and human experts: algorithms efficiently narrow the search space, while experts provide the scientific context that distinguishes meaningful anomalies from statistical curiosities.

\section{Methodology}

\subsection{Experimental Design}

This study evaluates four Isolation Forest-based anomaly-ranking strategies for identifying scientifically interesting radio sources: Standard Enhanced, Pooled Gain, Kurtosis-based, and Hypervolume. The four methods are evaluated using the same input feature representation, evaluation subset, ground-truth labels, and performance measures to ensure a controlled comparison.

\begin{figure}[htbp]
    \centering
    \includegraphics[width=\linewidth]{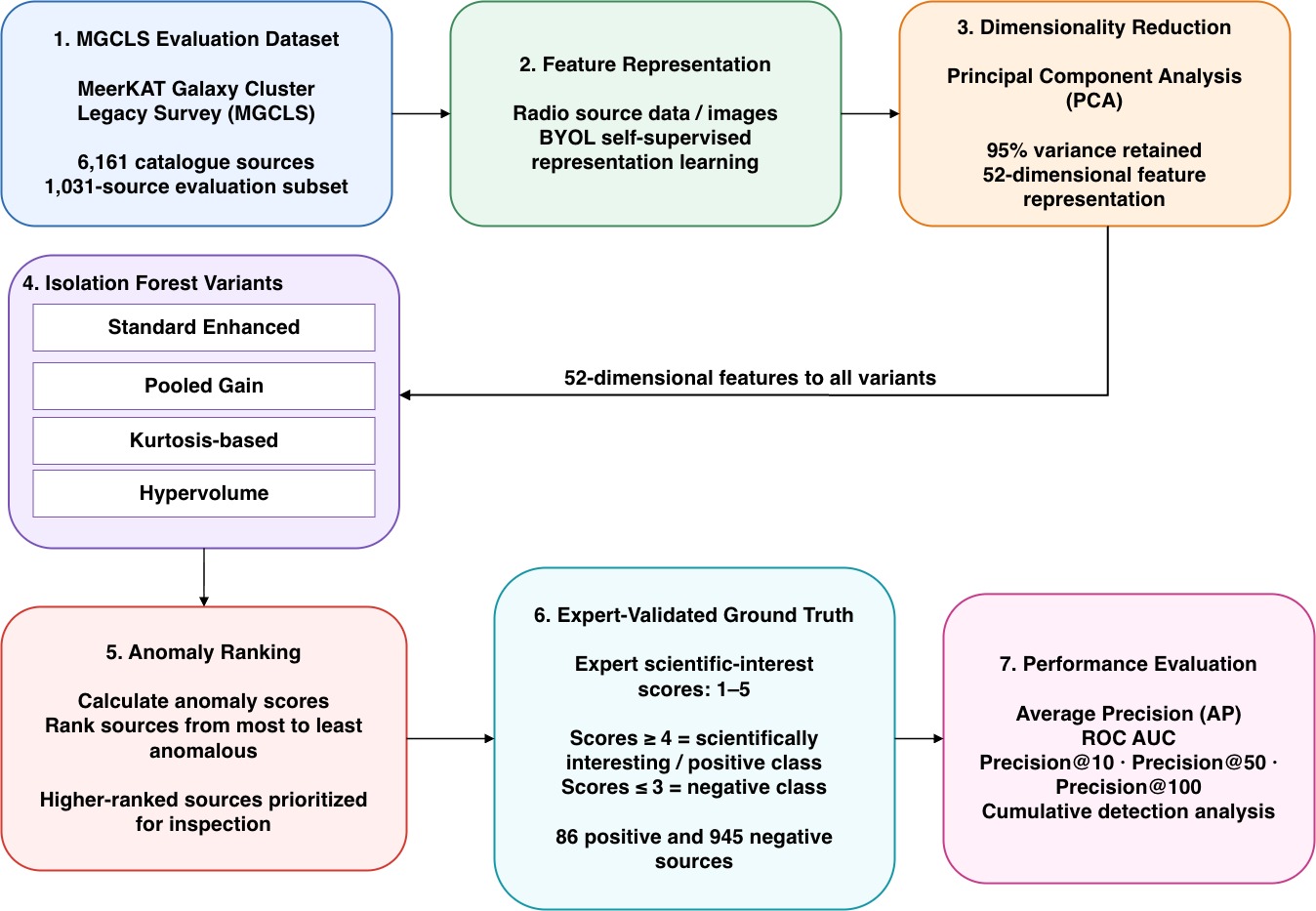}
    \caption{Overall experimental workflow for the comparative evaluation of the four Isolation Forest variants.}
    \label{fig:experimental_workflow}
\end{figure}

The overall experimental workflow is shown in Fig.~\ref{fig:experimental_workflow}. It consists of five stages: (i) selection of the MGCLS evaluation sources, (ii) construction of the 52-dimensional feature representation using BYOL and PCA, (iii) application of the four Isolation Forest variants, (iv) ranking of sources according to their anomaly scores, and (v) comparison of the resulting rankings against expert-validated scientific-interest labels. The evaluation focuses on early-stage ranking because the intended application is to assist astronomers in prioritizing a small number of candidates for expert inspection, while the evaluated extensions are motivated by non-uniform feature and split selection \cite{b24} and alternative distribution- and volume-based scoring \cite{b25}.
%MAYI
% The evaluation focuses particularly on early-stage ranking because the intended application is to assist astronomers in prioritizing a small number of candidates for expert inspection. This design is consistent with the human-in-the-loop philosophy of ASTRONOMALY and ASTRONOMALY Protege, where automated methods reduce the search space before expert feedback is applied \cite{b3,b4}. The underlying Isolation Forest methodology follows Liu et al. \cite{b23}, while the feature-aware extensions considered in this study are motivated by subsequent work on non-uniform feature and split selection \cite{b24} and alternative distribution- and volume-based scoring \cite{b25}.
% The evaluation focuses on early-stage ranking because the intended application is to assist astronomers in prioritizing a small number of candidates for expert inspection. The evaluated extensions are motivated by non-uniform feature and split selection \cite{b24} and alternative distribution- and volume-based scoring \cite{b25}.

\subsection{Dataset Description and Preprocessing}

The evaluation uses the MeerKAT Galaxy Cluster Legacy Survey (MGCLS), a radio survey conducted with the MeerKAT telescope. The MGCLS catalogue contains 6,161 sources, from which the 1,031-source evaluation subset used in the ASTRONOMALY Protege study was selected \cite{b2,b4}. Each source in the evaluation subset is represented by a 52-dimensional feature vector generated using the representation adopted by ASTRONOMALY Protege, obtained using Bootstrap Your Own Latent (BYOL) self-supervised learning \cite{b28} followed by principal component analysis (PCA), which retains 95 percent of the variance. Expert assessments provide scientific-interest scores ranging from 1 to 5; sources assigned scores of 4 or 5 are treated as scientifically interesting (positive class), while scores of 1, 2, or 3 form the negative class, consistent with the ASTRONOMALY Protege formulation \cite{b4}. The resulting evaluation set contains 86 positive and 945 negative sources, corresponding to an anomaly prevalence of 8.3 percent. Table~\ref{tab:dataset_composition} summarizes the composition of the evaluation dataset.

\begin{table}[htbp]
\centering
\caption{Dataset composition.}
\label{tab:dataset_composition}
\begin{tabular}{lrr}
\toprule
\textbf{Expert score} & \textbf{Number of sources} & \textbf{Percentage} \\
\midrule
5              & 24  & 2.3\%  \\
4              & 62  & 6.0\%  \\
3              & 122 & 11.8\% \\
2              & 439 & 42.6\% \\
1              & 384 & 37.2\% \\
\midrule
Positive ($\geq$4) & 86  & 8.3\%  \\
Negative ($\leq$3) & 945 & 91.7\% \\
\midrule
Total          & 1{,}031 & 100\% \\
\bottomrule
\end{tabular}
\end{table}

\subsection{Algorithm Implementation and Experimental Configuration}

% Four Isolation Forest variants were implemented and evaluated using a common 52-dimensional feature representation: Standard Enhanced, Pooled Gain, Kurtosis-based, and Hypervolume. The methodological principles underlying these approaches are described in Section II. In this study, the four variants were implemented within a common EnhancedIsolationForest framework to ensure that the comparison differed primarily in the feature-selection or anomaly-scoring strategy. 

All four variants used the common 52-dimensional feature representation described above and were implemented within a common EnhancedIsolationForest framework, ensuring that the comparison differed primarily in the feature-selection or anomaly-scoring strategy.
 
% The Standard Enhanced variant used random feature selection and a uniformly sampled split-value at each internal tree node. The Pooled Gain variant selected features probabilistically according to a combined importance measure based on feature variance and absolute kurtosis. Specifically, the implementation assigned weights of 0.7 to normalized variance and 0.3 to normalized absolute kurtosis. The resulting feature-importance values were normalized to form a probability distribution from which the splitting feature was sampled. This implementation was motivated by previous work on non-uniform feature selection in Isolation Forests \cite{b24}.

% The Kurtosis-based variant selected splitting features according to their normalized absolute kurtosis, thereby assigning greater selection probability to features with heavier-tailed distributions. The Hypervolume variant used a different anomaly-scoring procedure based on the terminal leaf reached by each source. For a source reaching a leaf containing $m$ samples, the implementation assigned a leaf-level score of
% {\small
% \begin{equation}
% v(x) = \frac{1}{m+1}.
% \end{equation}
% }
% The score for a source was obtained by averaging this quantity across all isolation trees. This leaf-size-based scoring strategy was evaluated as a volume-inspired alternative to conventional path-length scoring, consistent with the broader motivation of distribution- and volume-based Isolation Forest scoring \cite{b25}.

The Standard Enhanced variant used random feature selection and uniformly sampled split values. Pooled Gain and Kurtosis-based followed the feature-selection strategies described in Section II: Pooled Gain combined normalized variance and absolute kurtosis with weights of 0.7 and 0.3, respectively, while Kurtosis-based used normalized absolute kurtosis alone \cite{b24}. Hypervolume differed primarily in its anomaly-scoring procedure. For a source reaching a terminal leaf containing $m$ samples, the leaf-level score was defined as

{\small
\begin{equation}
v(x) = \frac{1}{m+1}.
\end{equation}
}

The final Hypervolume score was obtained by averaging this quantity across all isolation trees, providing a volume-inspired alternative to conventional path-length scoring \cite{b25}.

For the three depth-based variants, anomaly scores were calculated from the mean path length using the standard Isolation Forest formulation \cite{b23}. All variants used the common experimental configuration summarized in Table~\ref{tab:hyperparams}. A fixed random seed of 42 was used for the reported comparison; because only one seed was evaluated, the results do not establish robustness to stochastic variation.

\begin{table}[htbp]
\caption{Experimental configuration.}
\label{tab:hyperparams}
\centering
\begin{tabular}{lcl}
\toprule
\textbf{Parameter} & \textbf{Value} & \textbf{Configuration} \\
\midrule
Number of trees & 100 & Common to all variants \\
Subsample size & 256 & Maximum samples per tree \\
Maximum tree depth & 10 & Common tree-building limit \\
Random state & 42 & Fixed for reproducibility \\
Feature dimension & 52 & BYOL + PCA representation \\
\hline
\end{tabular}
\end{table}

\begin{figure}[htbp]
    \centering
    \includegraphics[width=\linewidth]{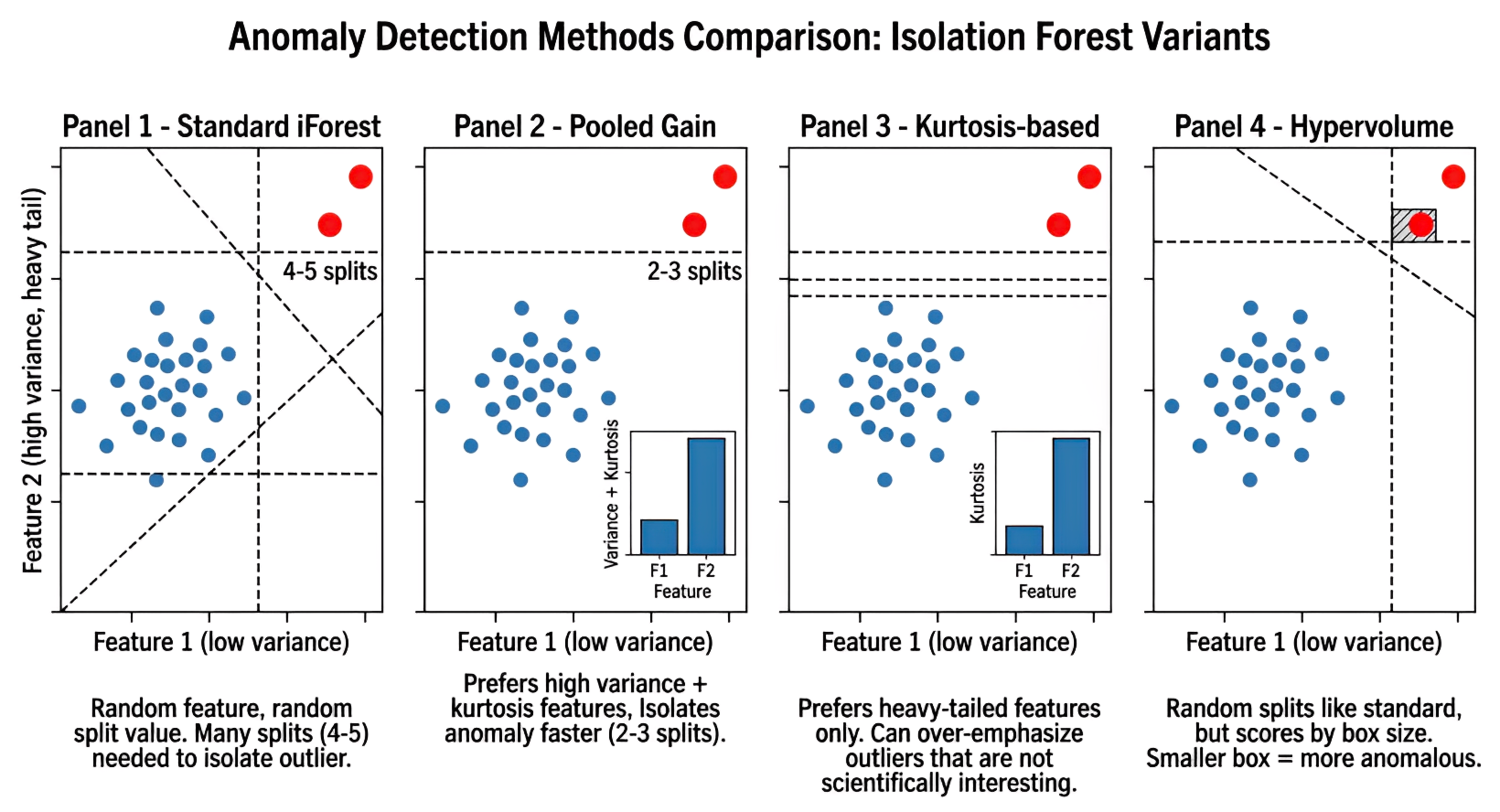}
    \caption{Comparison of the splitting and scoring strategies used by the four evaluated Isolation Forest variants.}
    \label{fig:splitting}
\end{figure}

Figure~\ref{fig:splitting} summarizes the principal splitting and scoring differences among the four evaluated variants.

\subsection{Evaluation Protocol}

The four variants were evaluated using the same expert-validated binary labels and ranking procedure. For each method, the model was fitted to the 1,031-source evaluation subset and an anomaly score was calculated for every source. The resulting scores were used to construct Precision--Recall and receiver operating characteristic (ROC) curves and ranked source lists.

Average Precision (AP) summarizes performance over the Precision--Recall curve and is appropriate for the imbalanced positive class (8.3\%). AP was calculated as
{\small
\begin{equation}
AP = \sum_{k} P(k)\Delta R(k),
\end{equation}
}
where $P(k)$ is the precision at a given rank and $\Delta R(k)$ is the change in recall.

The ROC curve was constructed using the true-positive rate (TPR) and false-positive rate (FPR):
{\small
\begin{equation}
TPR = \frac{TP}{TP+FN},
\end{equation}
}
and
{\small
\begin{equation}
FPR = \frac{FP}{FP+TN}.
\end{equation}
}
The area under the ROC curve (ROC AUC) was then calculated as
{\small
\begin{equation}
AUC = \int_{0}^{1} TPR(FPR)\,dFPR.
\end{equation}
}

Because the intended application is to prioritize a small number of sources for expert inspection, Precision@K was calculated at $K=10$, $50$, and $100$:
{\small
\begin{equation}
P@K = \frac{1}{K}\sum_{i=1}^{K} y_{(i)},
\end{equation}
}
where $y_{(i)}$ denotes the binary expert-validated label of the source at rank $i$. Thus, $P@10$, $P@50$, and $P@100$ represent the proportion of expert-validated anomalies among the first 10, 50, and 100 ranked sources.

Cumulative anomaly detection was calculated by ordering sources according to anomaly score and computing the cumulative number of positive labels:
{\small
\begin{equation}
CD(K) = \sum_{i=1}^{K} y_{(i)}.
\end{equation}
}
This metric indicates how many expert-validated scientifically interesting sources are identified within the first $K$ ranked sources.

To ensure a consistent ranking direction, scores from the three depth-based variants were sign-reversed because lower raw scores correspond to greater anomaly likelihood. Hypervolume scores were used directly because higher values indicate greater anomaly likelihood.

\section{Results}

\subsection{Precision-Recall Analysis}

\begin{figure}[htbp]
\centerline{\includegraphics[width=0.9\linewidth]{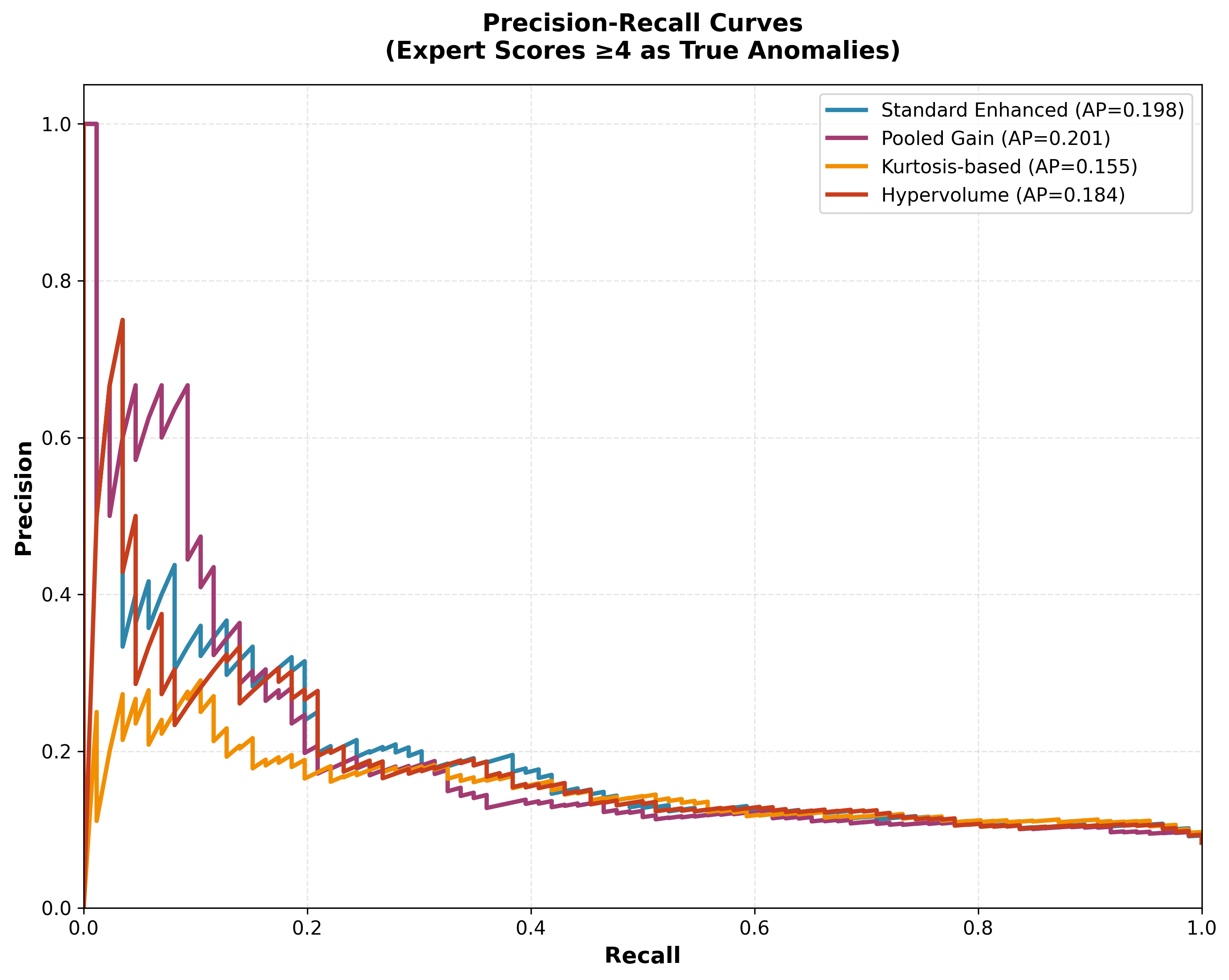}}
\caption{Precision--Recall curves for the four Isolation Forest variants, with expert scores $\geq 4$ treated as the positive class.}
\label{fig:pr}
\end{figure}

The Pooled Gain variant achieved the highest Average Precision
(AP = 0.201), followed by Standard Enhanced (0.198),
Hypervolume (0.184), and Kurtosis-based (0.155). The higher AP
obtained by Pooled Gain indicates better average precision across the
Precision--Recall curve, which is particularly relevant given the
imbalanced positive class (8.3\%).

\subsection{ROC Curve Analysis}

\begin{figure}[htbp]
\centerline{\includegraphics[width=0.9\linewidth]{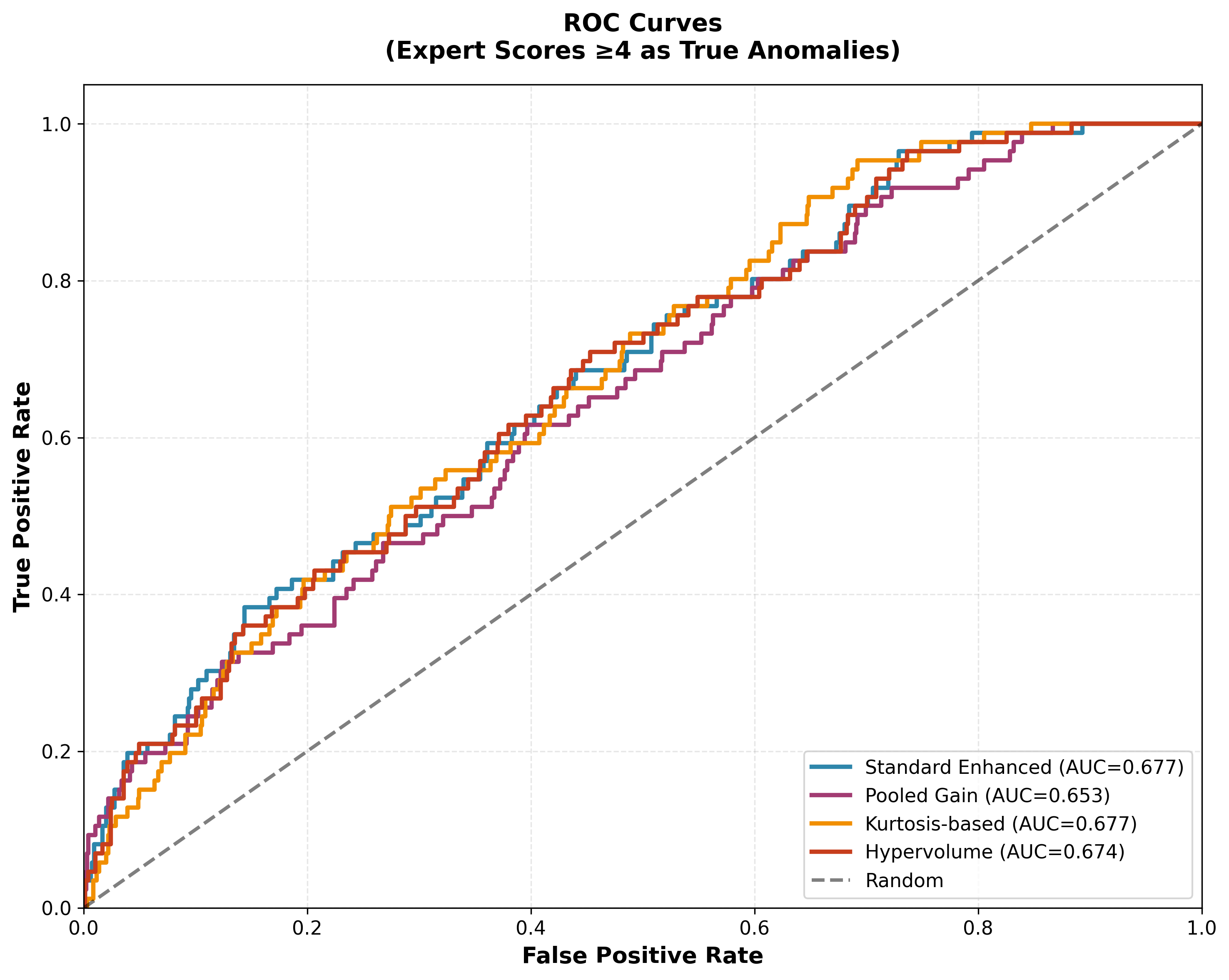}}
\caption{ROC curves for the four Isolation Forest variants, with expert scores $\geq 4$ treated as the positive class.}
\label{fig:roc}
\end{figure}

Figure~\ref{fig:roc} presents the ROC curves for the four Isolation
Forest variants.

Standard Enhanced and Kurtosis-based achieved the highest ROC AUC
(0.677), followed by Hypervolume (0.674) and Pooled Gain (0.653).
All four variants remained above the random-classifier baseline of
0.5. Although Pooled Gain has the lowest ROC AUC, its higher Average
Precision and Precision@10 indicate stronger early-stage ranking
performance.

\subsection{Cumulative Detection Analysis}

\begin{figure}[htbp]
\centerline{\includegraphics[width=0.9\linewidth]{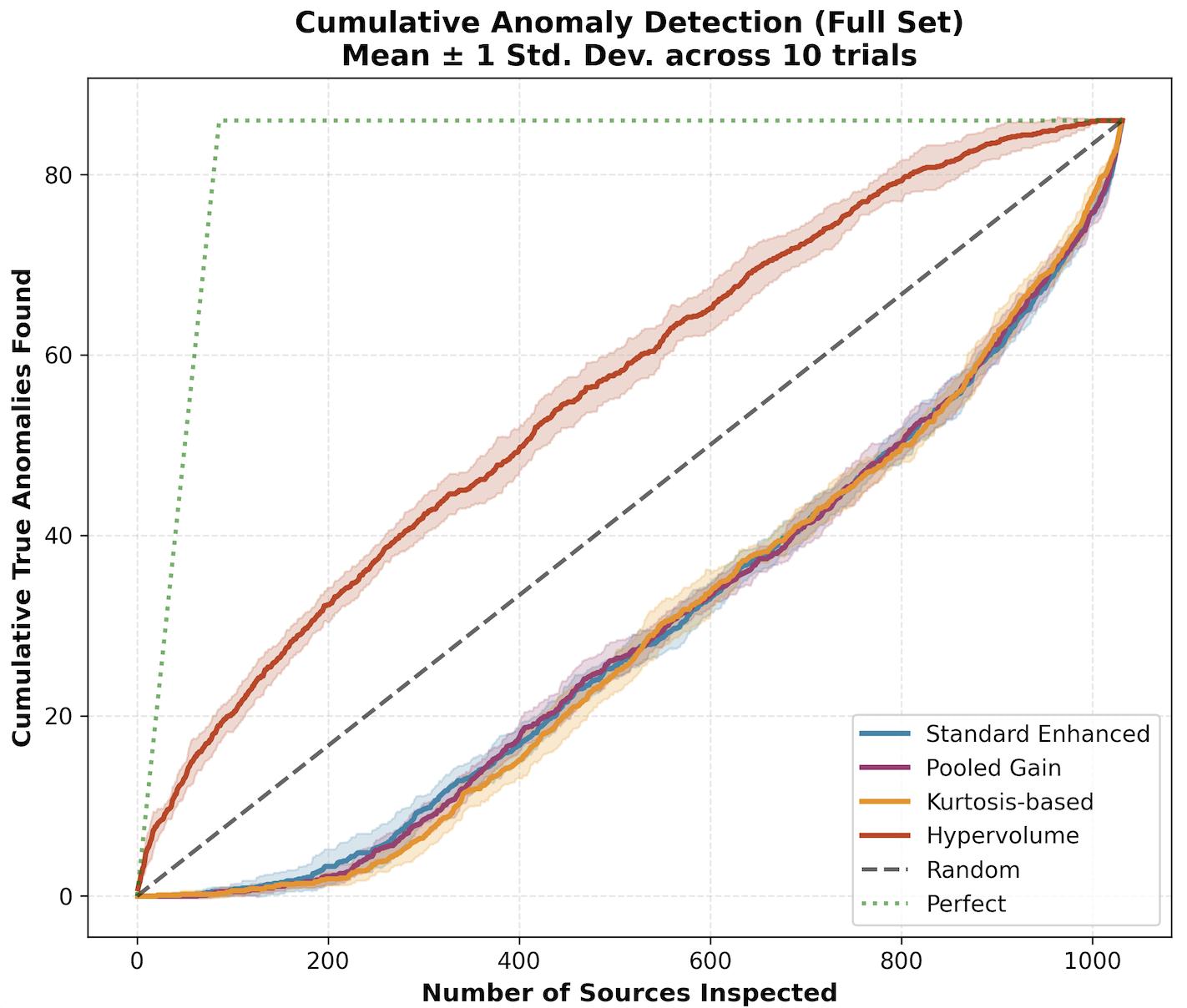}}
\caption{Cumulative detection of expert-validated scientifically interesting sources for the full evaluation set.}
\label{fig:cumulative_full}
\end{figure}

\begin{figure}[htbp]
\centerline{\includegraphics[width=0.9\linewidth]{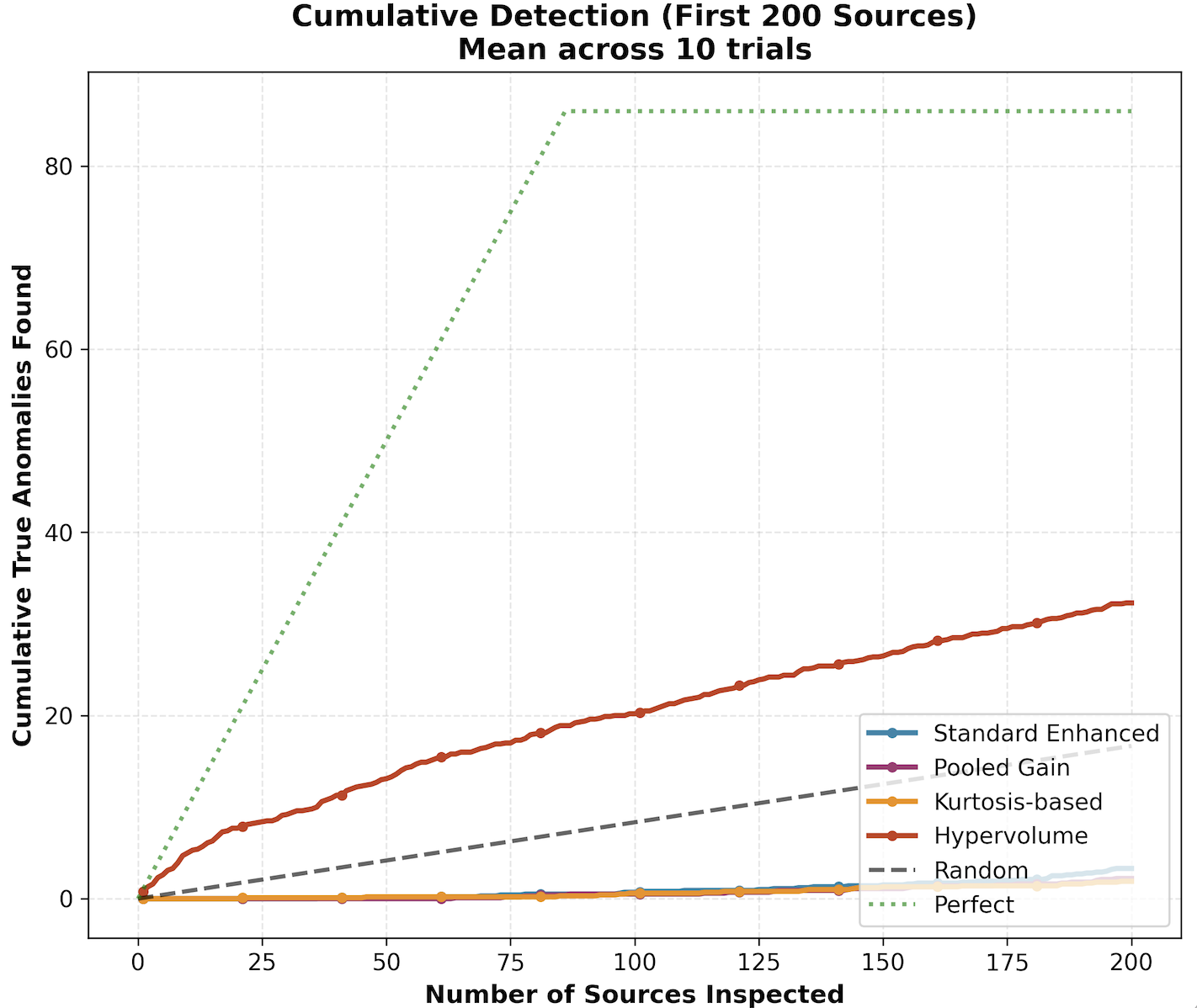}}
\caption{Cumulative detection of expert-validated scientifically interesting sources among the first 200 ranked candidates.}
\label{fig:cumulative_first200}
\end{figure}

Figures~\ref{fig:cumulative_full} and
\ref{fig:cumulative_first200} present the cumulative detection curves
for the full set of sources and the first 200 sources, respectively.
These plots address the practical question of how many expert-validated
scientifically interesting sources an astronomer would encounter when
inspecting the top $N$ ranked candidates.

Pooled Gain demonstrates the strongest early-stage performance, with
P@10 = 0.600, meaning that 6 of the top 10 ranked sources belonged to
the positive class. Given the overall positive-class prevalence of
8.3\%, a random selection of 10 sources would be expected to contain
approximately 0.83 positive sources. The observed six positive sources
therefore correspond to an enrichment of approximately 7.2-fold over
random selection.

At deeper inspection levels, performance differences become less
pronounced, with Standard Enhanced leading at both P@50 and P@100.
The relative ordering of the methods therefore varies with the number
of sources inspected.

\subsection{Performance Summary}

Table~\ref{tab:performance} summarizes the performance metrics for all
four variants.

\begin{table}[htbp]
\caption{Performance comparison of the four Isolation Forest variants using expert-validated labels.}
\label{tab:performance}
\centering
\begin{tabular}{lccccc}
\toprule
\textbf{Algorithm} & \textbf{AP} & \textbf{AUC} & \textbf{P@10} & \textbf{P@50} & \textbf{P@100} \\
\midrule
Standard Enhanced & 0.198 & 0.677 & 0.400 & 0.320 & 0.210 \\
Pooled Gain & 0.201 & 0.653 & 0.600 & 0.280 & 0.180 \\
Kurtosis-based & 0.155 & 0.677 & 0.200 & 0.220 & 0.170 \\
Hypervolume & 0.184 & 0.674 & 0.400 & 0.300 & 0.200 \\
\bottomrule
\end{tabular}
\end{table}

% Pooled Gain achieves the highest Average Precision (0.201) and
% Precision@10 (0.600), indicating the strongest early-stage ranking
% performance among the evaluated variants. Standard Enhanced achieves
% the highest ROC AUC (0.677, tied with Kurtosis-based) and the highest
% Precision@50 (0.320) and Precision@100 (0.210). These results indicate
% that no single variant dominates across all evaluation criteria.
% Instead, Pooled Gain provides an advantage when the primary objective
% is to prioritize a small number of candidates for early expert
% inspection, whereas Standard Enhanced performs better at the broader
% top-50 and top-100 ranking levels.

Pooled Gain achieves the highest Average Precision (0.201) and Precision@10 (0.600), indicating the strongest early-stage ranking performance. Standard Enhanced and Kurtosis-based achieve the highest ROC AUC (0.677), while Standard Enhanced also leads at Precision@50 (0.320) and Precision@100 (0.210). These results show that no single variant dominates across all evaluation criteria: Pooled Gain is stronger for early candidate prioritization, whereas Standard Enhanced performs better at broader inspection levels.

\section{Discussion}

\subsection{Key Findings}

Three main insights emerge from this study:

\textbf{1. Feature-aware splitting improves early-stage ranking.}
The Pooled Gain variant achieved a 50\% relative improvement in P@10 over Standard Enhanced (0.600 vs. 0.400), indicating stronger
prioritization of expert-validated scientifically interesting sources among the highest-ranked candidates. This suggests that combining
variance and kurtosis information for feature selection can improve early-stage candidate ranking, although the effect should be validated across additional random seeds and datasets.

\textbf{2. Not all enhancements are equally effective.}
The Kurtosis-based variant produced the lowest Average Precision
(AP = 0.155) and Precision@10 (P@10 = 0.200), while Pooled Gain,
which combines variance and kurtosis information, achieved the highest
AP and P@10. This suggests that combining multiple feature
characteristics may be more effective than relying exclusively on
absolute kurtosis in this experimental setting.

\textbf{3. Unsupervised ranking has inherent limitations.}
The relatively modest Average Precision values (AP = 0.155--0.201)
indicate that statistical anomaly ranking does not fully capture the
expert-defined notion of scientific interest. This reinforces the
potential value of human-in-the-loop approaches, where automated
ranking is used to prioritize candidates for subsequent expert
assessment \cite{b3,b4}.

\subsection{Practical Implications for Survey Operations}

The results have potential operational significance for large-scale
radio surveys in which astronomers cannot inspect every detected source
individually. The stronger early-stage ranking achieved by Pooled Gain
suggests that feature-aware ranking may help concentrate scientifically
interesting candidates earlier in the inspection sequence, supporting
candidate prioritization in future facilities such as the SKA and human-in-the-loop frameworks such as ASTRONOMALY Protege \cite{b1,b4}. However, the extent to which this reduces actual expert
workload requires empirical validation.

\subsection{Limitations and Future Directions}

This study has several limitations. The experiments used a single
random seed (42), one 52-dimensional BYOL+PCA feature representation,
and a single MGCLS evaluation subset. In addition, the expert-derived
scientific-interest labels do not capture variability across multiple
independent experts. Future work should therefore evaluate the methods
across multiple random seeds, alternative feature representations and
radio surveys, incorporate multi-expert annotations, and empirically
evaluate integration with human-in-the-loop systems.

\section{Conclusion}
This study evaluated four Isolation Forest variants for ranking scientifically interesting radio sources using expert-validated labels from an MGCLS evaluation subset. Pooled Gain achieved the highest Average Precision (AP = 0.201) and Precision@10 (P@10 = 0.600), demonstrating the strongest early-stage candidate prioritization, while Standard Enhanced and Kurtosis-based achieved the highest ROC AUC (0.677), with Standard Enhanced also performing best at broader P@50 and P@100 inspection levels. These findings show that the most suitable Isolation Forest strategy depends on whether the priority is early candidate retrieval or broader discrimination performance. Future work should evaluate the methods across multiple random seeds, alternative feature representations, additional radio surveys, and multi-expert annotations, and should investigate integration with human-in-the-loop frameworks such as ASTRONOMALY Protege to determine whether improved early-stage ranking translates into reduced expert inspection effort and improved scientific discovery.

\end{document}